# Small-misorientation toughness in biominerals evolved convergently

Andrew J. Lew[1,2], Cayla A. Stifler[3], Connor A. Schmidt[3,4], Markus J. Buehler[1*], Pupa U. P. A. Gilbert[3,4,5,6, *]

[1] Laboratory for Atomistic and Molecular Mechanics (LAMM), Massachusetts Institute of Technology, 77 Massachusetts Ave., Cambridge, MA 02139, USA

[2] Department of Chemistry, Massachusetts Institute of Technology, 77 Massachusetts Ave., Cambridge, MA 02139, USA

[3] Department of Physics, University of Wisconsin, Madison, WI 53706, USA
[4] Department of Chemistry, University of Wisconsin, Madison, WI 53706, USA
[5] Departments of Materials Science and Engineering, Geoscience, University of Wisconsin, Madison, WI 53706, USA
[6] Chemical Sciences Division, Lawrence Berkeley National Laboratory, Berkeley, CA 94720, USA.

[*] Co-corresponding authors:
Email: mbuehler@mit.edu , pupa@physics.wisc.edu

## Abstract

The hardest materials in living organisms are biologically grown crystalline minerals, or biominerals, which are also incredibly fracture-tough. Biomineral "mesostructure" includes size, shape, spatial arrangement, and crystal orientation of crystallites, observable at the mesoscale (10 nm-10 μm). Here we show that diverse biominerals, including nacre and prisms from mollusk shells, coral skeletons, and tunicate spicules have different mesostructures, but they converged to similar, small (<30°) misorientations of adjacent crystals at the mesoscale. We show that such small misorientations are an effective toughening mechanism. Combining Polarization-dependent Imaging Contrast (PIC) mapping of mesostructures and Molecular Dynamics (MD) simulations of misoriented bicrystals we reveal here that small misorientations toughen bicrystals, thus explaining why they evolved independently but convergently: preventing fracture is a clear evolutionary advantage for diverse organisms.

# Introduction

Biominerals are crystalline composites formed by living organisms [1, 2, 3, 4, 5, 6]. Diverse phyla, such as cnidarians, ascidian tunicates, and mollusks diverged from one another 600-800 million years ago (Ma) [7], but started making their biominerals during the Phanerozoic (541 Ma-present), with the oldest fossils for these three phyla being 510, 270, and 535 Ma, respectively [8, 9, 10]. Thus, if any biomineral structure or formation mechanism is observed in multiple biominerals, these must have evolved independently, or "convergently", as termed in biology for bat and bird wings similarly structured for flying, or fish, dolphins, and diving birds with body plans streamlined for swimming. The three above phyla were recently shown to make their biominerals convergently [7], in that they all form by particle attachment [11] of amorphous calcium carbonate precursors [7]. Is it possible that (mis)orientation of immediately adjacent crystals at the nano- or micro-scale is also convergent across phyla? And if so, why? In other words, what could the evolutionary advantage of (mis)orientation found in diverse biominerals be? Previous work using electron backscatter diffraction (EBSD) [12, 13, 14], or x-ray diffraction (XRD) [15, 16], hinted at the possibility that adjacent crystals are only slightly misoriented (<30°), but a direct quantitative comparison across biominerals from diverse phyla remains elusive.

The lack of comprehensive orientation information across biominerals, thus far, can be attributed to several factors. One is that sensitivity to $a$-, $b$-, and $c$-axis orientation, shared by EBSD and XRD, hinders such analysis because of extensive crystal twinning in the $ab$-plane [15, 17], which is not relevant to materials properties. Other factors include the complex arrangement of too many biomineral crystals analyzed at once with penetrating XRD, and/or insufficient resolution with surface-sensitive EBSD.

All biominerals exhibit a complex arrangement of crystallites with 10 nm-10 µm sizes [18], diverse shapes, spatial arrangements, and crystal misorientations with respect to one another, collectively defined here as mesoscale (10 nm-10 µm) structure or simply "mesostructure". Each biomineral has a characteristic mesostructure, distinct from others in different animal genera.

Here we analyzed diverse biominerals with Polarization-dependent Imaging Contrast (PIC) mapping, which effectively and quantitatively reveals every aspect of biomineral mesostructure, combining surface sensitivity (5 nm at the O K-edge, 3 nm at the Ca L-edge [19]), high spatial resolution (<60 nm, down to 10 nm pixels [20, 21]), and large-area imaging [22].

# Results

The skeletons of stony corals, the spicules of tunicate ascidians, and the prisms of bivalve mollusk shells provide structural support or sheltering to the animal that builds them. These three biominerals are made of one of the three calcium carbonate ($CaCO_3$) polymorphs: aragonite, vaterite, or calcite, respectively. The mesostructures of three representative biominerals from these diverse animals are presented in the PIC maps of Fig. 1.

In a PIC map the color represents quantitatively the in-plane and off-plane orientation of the *c*-axis of the crystal in each nano-scale pixel, as displayed by hue and brightness, respectively [20]. The PIC maps in Fig. 1, with pixels size 56 nm and probing depth 5 nm at oxygen K-edge [23], reveal diverse mesostructures for the three biominerals: aragonite coral skeletons are spherulitic, that is, acicular crystals (fibers F) have their *c*-axes radiating from common centers termed centers of calcification (CoCs). The CoCs form tortuous paths, and the fibers do not form spheres but bundles of crystals that can grow in any direction and stop growing when they run into other bundles. Within each bundle, however, adjacent fiber crystals are nearly cooriented with misorientation of adjacent crystals never exceeding ~30° [21, 22]. In Fig.1A this quantitative result is visible at a glance: adjacent fibers are either the same color or two adjacent colors, such as green and yellow, or blue and cyan. In tunicate ascidian vaterite spicules a similar result is observed: the left spicule in Fig.1B is cyan and green, the central one in cross-section is green and yellow, and the right one is magenta, with spines that are red on one side or blue on the other side. The calcite prisms of most bivalve mollusk shells are single crystalline [24]. *Pinctada fucata* and *Pinctada margaritifera* prisms, however, show domains of aggregated particles with different orientations [4, 16, 20, 25]. The *Pinctada margaritifera* prisms in Fig.1C show the most commonly observed case with nearly misoriented crystals, all within the red-orange range of colors. Even in the rare prisms with a greater diversity of orientations, such as those in Fig.1D, adjacent crystals have adjacent colors.

Even in the most iconic biomineral, abalone nacre, adjacent crystalline tablets are closely misoriented, with *c*-axis misorientation angles within 30° (± 15° from the normal to nacre layers) [26, 27], which are abiotically selected by a competition for space growth model [13, 28, 29]. What had not been previously observed, however, is that the layer of spherulites that occasionally grow within nacre [30], seed the crystal orientations for nacre tablets, and then competition for space selects for *c*-axes to be oriented ± 15° from the normal to nacre tablet layers. In Fig.2 this behavior can be observed directly: steady-state nacre is all cyan (0°), or bluish cyan (-15°) or greenish cyan (+15°) in PIC maps, as seen at the top and bottom of Fig.2. The layer of spherulites in Fig.2 clearly has only small misorientation of adjacent acicular crystals, as observed in all spherulites, synthetic and biogenic [21], and in the coral skeleton in Fig.1A. The spherulitic crystals seed nacre tablet crystal orientation, become layered like nacre, and then, because more misoriented crystals grow more slowly [28], the fastest growing ones with *c*-axes ± 15° from the normal fill space and thus prevail over the blue or green crystal orientations near the center of Fig.2.

Having observed small misorientations in four diverse biominerals, the question is why did these evolve convergently, despite divergent mesostructure, to have similar misorientation of each two adjacent crystals? To address this question, we carried out molecular dynamics (MD) simulations of bicrystals, in which the *c*-axes of the two crystals are either cooriented ($\theta = 0°$) or misoriented by 10°-90°. The bottom crystal is notched, and a tensile load is applied horizontally to induce bicrystal fracture, starting from the notch and continuing through or being deflected by the bicrystal interface (mode I fracture)(Fig.3A). All three polymorphs were used, and each simulation was repeated three times. The results are presented in Fig.3B and clearly show crack deflection, compared to the single crystal (0°), at a small misorientation angle of 10°, which becomes a less distinct effect when the misorientation is increased to a larger 45° angle. Fracture toughness as a function of misorientation, measured as the integral under the stress-strain curves (Fig.S1), peaks at small angles of 10°, 20°, and 30° for aragonite, vaterite, and calcite, respectively, as shown in Fig.3C.

In the case of small misorientations, increased crack deflection across the interface correlates with increase in fracture toughness. However, at large angles aragonite exhibits another increase in toughness distinct from the small misorientation effect. This is likely due to the anisotropic fracture properties of single crystal aragonite having higher toughness when pulled perpendicular to the *c*-axis than parallel to it (horizontally), as shown in Fig.S2B. Additionally, vaterite single crystals exhibit secondary fracture from the upper unnotched plane when *c*-axes are rotated to large angles away from the horizontal tensile direction, as shown in Fig.S2C, resulting in crack paths that appear tortuous without true misorientation toughening. Similarly, calcite single crystals fail much more readily as their *c*-axes are rotated away from the horizontal tensile direction, with lower yield strains as shown in Fig.S2D, acting as a source of secondary fracture and providing spuriously torturous fracture paths without correspondingly high toughness. All three of these large-angle behaviors are due to fracture properties of single crystals, not due to large misorientations of bicrystals.

Despite these polymorph-specific single crystal behaviors, the connecting trend is clear: small misorientation makes bicrystals tougher. By how much are these misoriented bicrystals more fracture-tough than cooriented crystals? Direct comparison of the MD results enables a precise answer: twice, 2.5, and 3 times tougher for aragonite, vaterite, and calcite, respectively. This comparison is presented in Fig.4, which includes the result obtained for hydroxyapatite in human enamel using similar MD simulations [31].

## Discussion

No two biomineral mesostructures could more diverse than nacre and spherulitic coral skeletons, yet from the point of view of small-misorientation of adjacent crystals, nacre and corals skeletons are quite similar. One could even say that *nacre is spherulitic*. In fact, nacre can start spherulitic, as shown in Fig.2, and then small misorientations are selected abiotically, in competition for space.

As an aside, many blocky aragonite crystals in the literature are incorrectly termed spherulitic [28], because they are blocky and single crystalline, not a distribution of acicular crystals, slightly and gradually changing in orientation like the spherulites in Figs.1A, 2. The observation that spherulitic aragonite seeds the crystal orientation and growth of nacre after an organic layer, to our knowledge, was only done in this work.

Many toughening mechanisms have been observed in biological materials [32, 33]. Small misorientation (<30°) of adjacent crystals is an additional toughening mechanism as shown by the significant increase in toughness compared to cooriented bicrystals (Figs.3, 4).

Gradually changing small misorientations, as observed in spherulites, coral skeletons, and nacre are effectively gradient orientations. Many materials properties have been observed to vary gradually, and thus have functional gradients [34]. These include composition, dimensions, arrangement, orientation of fibers, distribution, and interfaces [34]. Gradient crystal orientations must be added to the list of functional gradients. They are an important protagonist to the evolution of animal life, so much so that they are observed in animals that diverged from one another before they started making small misorientations, in fact, before they started making any biominerals at all. In other words, small misorientation is convergent.

Small misorientations were recently observed in human enamel, and shown, with MD simulations similar to those presented here, to be functional in preventing fracture in human enamel [31]. The fact that enamel, different in mineralogy and mesostructure from $CaCO_3$ biominerals, also converged to similar small misorientations further strengthens the conclusion that they are functional, toughen the biominerals, and thus provide evolutionary advantages to the organisms that form them.

## Funding Sources

PG acknowledges 50% support from the U.S. Department of Energy, Office of Science, Office of Basic Energy Sciences, Chemical Sciences, Geosciences, and Biosciences Division, under Award DE-FG02-07ER15899, 40% support from the Laboratory Directed Research and Development (LDRD) program at Berkeley Lab, through DOE-BES, under Award Number DE-AC02-05CH11231, and 10% support from NSF grant DMR-1603192. PEEM experiments were done at the Advanced Light Source (ALS), which is supported by the Director, Office of Science, Office of Basic Energy Sciences, US Department of Energy under Contract No. DE-AC02-05CH11231. AJL acknowledges support by the NSF GRFP under Grant No. 1122374. AL acknowledges support from NSF GRFP under Grant No. 1122374. MJB and AJL acknowledge support by the Office of Naval Research (N000141612333 and N000141912375), AFOSR-MURI (FA9550-15-1-0514) and the Army Research Office (W911NF1920098).

## Acknowledgements

We thank Andrew H. Knoll for discussions, Tali Mass for providing the *Stylophora pistillata* coral skeleton, Noa Shenkar for providing the tunicate ascidians *Herdmania momus* from the Steinhardt Museum of Natural History, Tel-Aviv, Israel. We are grateful to Nobumichi Tamura for providing precise xyz coordinates of atoms, unit cell dimensions, and angles for aragonite, vaterite, and calcite for the MD simulations, and to Rajesh V. Chopdekar and Roland Koch for technical assistance during beamtime on PEEM-3 at ALS.

## Author contributions

PG conceived the idea that small misorientations are convergent for a materials reason, and that MD simulations could be used to test this hypothesis. MJB and AJL designed the MD simulations and fracture analyses, and AJL carried out the atomistic-level simulations, collected the data, and visualized the results. The simulations were analyzed and interpreted (in conjunction with the experimental data) by AJL, MJB and PG. CAS, CAS, and PG acquired all the PIC mapping data. PG wrote the manuscript, all co-authors edited it.

## Methods

### Samples

*Stylophora pistillata* coral skeletons (Fig.1A) were provided by Tali Mass, and they originated from the Red Sea near Eilat, Israel.

*Herdmania momus* tunicate ascidians (Fig.1B) were collected in 2004 in Eilat, Israel, at a depth of 15 m, and preserved in ethanol the Steinhardt Museum of Natural History, Tel-Aviv, Israel, and were provided to us by Noa Shenkar. In August 2019 we bleached entire tunicate ascidians for 2 weeks in covered Petri dishes, then collected the vaterite spicules at the bottom of the dish, rinsed them in ethanol twice, and mounted, under a stereomicroscope, onto embedding molds.

*Pinctada margaritifera* (Fig.1C) were collected in 2004 from the Ferme Perlière Paul Gauguin, Rangiroa, French Polynesia. In 2021, a 1 cm x 1 cm portions of one shell valve were cut with a diamond saw, and mounted vertically onto embedding molds, to expose the shell cross-section for analysis in PEEM.

*Haliotis rufescens* California red abalone (Fig.2) shells were acquired from Monterey Abalone Company (Monterey, CA, USA). In 2020, SEM experiments revealed that one shell had perfectly formed spherulites. In 2021, a 1 cm x 1 cm portions of one shell were cut with a diamond saw, and mounted vertically onto embedding molds, to expose the shell cross-section for analysis in PEEM.

The samples from all 4 biominerals were embedded into EpoFix (EMS, Hatfield, PA, USA) and polished them with an AutoMet 250 Pro Grinder Polisher (Buehler, Lake Bluff, IL, USA), and analyzed with Polarized Light Microscopy (PLM). The best samples (e.g., the abalone shell that best showed spherulites in PLM) was identified using PLM and selected for PEEM analysis. The 4 selected samples were coated with 1 nm Pt in the region of interest, and 40 nm Pt surround it [35]. Since at the O K-edge the maximum probing depth is 5 nm [23], 1 nm guaranteed conductivity and therefore stability at high voltage, but no significant attenuation of the signal from the underlying biomineral surface. Forty nm Pt surround the region of interest guaranteed a strong and stable electrical contact with the sample holder, which is floating at -18 kV during the PEEM experiment.

All sample preparations were done in the Gilbert Group (GG) Lab at UW-Madison, WI, USA.

## PhotoEmission Electron Microscopy (PEEM)

PEEM experiments were done as described in [19], using the PEEM-3 instrument on beamline 11.0.1.1 at the Advanced Light Source, Lawrence Berkeley National Laboratory, Berkeley, CA, USA. The sample surface was mounted vertically, and the beam had a 60° incident angle (from the normal to the sample surface, which is also the PEEM optical axis), from the right.

## PIC or Polarization-dependent Imaging Contrast Mapping

For PIC mapping [20], stacks of PEEM images were acquired at the oxygen K-edge carbonate π* peak energy of 534 eV, with linear polarization from the elliptically polarizing undulator (EPU) varying from horizontal to vertical in 5° steps (19 images/stack). This rotation occurs in the polarization plane, to which all PIC mapping angles are referred: the in-plane angle, represented by hue, and the off-plane angle, represented by brightness in PIC maps. Black pixels have no polarization dependence, either because the material is not crystalline and dichroic, e.g., organic envelopes between calcite prisms in Fig.1C, CoCs in Fig.1A, large horizontal band of organic layer in Fig.2, or embedding epoxy around vaterite spicules in Fig.1B, or because the crystalline *c*-axis is pointing directly into the beam. This is the case for a few crystals on the right top third of the image in Fig.1A, or the one at bottom right in Fig.1D.

## MD Simulations

To model the effect of misorientation on the fracture behavior of aragonite, vaterite, and calcite, we represented the structure around grain boundaries as bicrystal samples consisting of two crystals stacked in the vertical direction and subjected them to tensile loading in the horizontal direction. The top crystal had lattice coordinates rotated about the depth axis to the desired misorientation angle θ while the bottom crystal's orientation was kept fixed with the crystalline *c*-axis horizontal, parallel to the loading direction [36]. A triangular notch 2 nm wide by 7 nm tall was carved into the bottom face to serve as the initial crack. This 0.8 nm thick bottom face was constrained to forbid vertical motion in order to enforce purely Mode I-fracture and eliminate shearing, which could significantly impact crack dynamics [37]. This sample setup is illustrated in Fig.3A. Single-crystal fracture was similarly conducted as a comparative control, with the same setting but with only one crystal, with its *c*-axis oriented to an angle θ' from the horizontal, as shown in Fig.S2A.

Simulations were run using the classical molecular dynamics (MD) code Large-scale Atomic/Molecular Massively Parallel Simulator (LAMMPS) [38]. Crystal equilibration and sintering were conducted with periodic boundary conditions. Mode I-fracture tensile tests were conducted with periodicity only in the depth-wise direction, with non-periodic conditions applied in the loading and crack propagation directions to remove the effects of lattice mismatch at the boundary and probe the effects of a single surface notch rather than infinite internal notches [36, 39]. Established force field parameters developed by Raiteri et al. [40, 41] applicable to $CaCO_3$ were used. However, long range Coulombic solvers responsible for computing charge interactions in an infinite array of periodic images [38] were omitted for proper treatment of non-periodic systems.

Crystal coordinates for aragonite [42], vaterite [43], and calcite [44] were first equilibrated to 300 K and 1 bar. From equilibrated single crystal structures, misorientation grain boundaries were formed by constructing two grains with the desired orientations and sintering them together. This sintering process consisted of four stages [31]: increasing pressure in the vertical direction from 1 bar to 10 GPa over 10 ps to press the two crystals together, holding the pressure at 10 GPa for 10 ps, decreasing the pressure down to 1 bar over 10 ps, and then holding at 1 bar for 10 ps to allow relaxation of residual stresses. This four-stage sintering process was repeated three times to ensure a well-formed interface. Misorientations of 5°, 10°, 20°, 30°, 45°, 60°, 75°, and 90° were investigated, and Mode I-fracture of each misorientation case was run three times in order to account for differences from random thermal fluctuations.

To facilitate tensile fracture, 0.8 nm thick walls were pulled away from each other at a 10 m/s strain rate in a two-part process consisting of deformation and equilibration steps, repeated until fracture completion [39, 45]. Specifically, we alternated between 0.5 ps of deformation with NVE ensemble to ensure energy conservation, and 1 ps of fixed sample length equilibration at 300 K with NVT ensemble, until a total strain of 10%.

### Fracture Visualization

Visualization of atomic structure and fracture dynamics was done with Open Visualization Tool (OVITO) [46]. To aid in simplifying the visualization of crack paths, atomic positions were frozen to their initial positions and represented in greyscale. Then, fracture path images were obtained by coloring atoms according to the percentage of neighbors lost over the course of fracture - white corresponding to unperturbed crystal structure and black corresponding to atoms that have lost half of their neighbors (*i. e.* at newly created fracture surfaces). In this way, we could visualize the precise path along which the crystal structure failed.

# Figures and captions

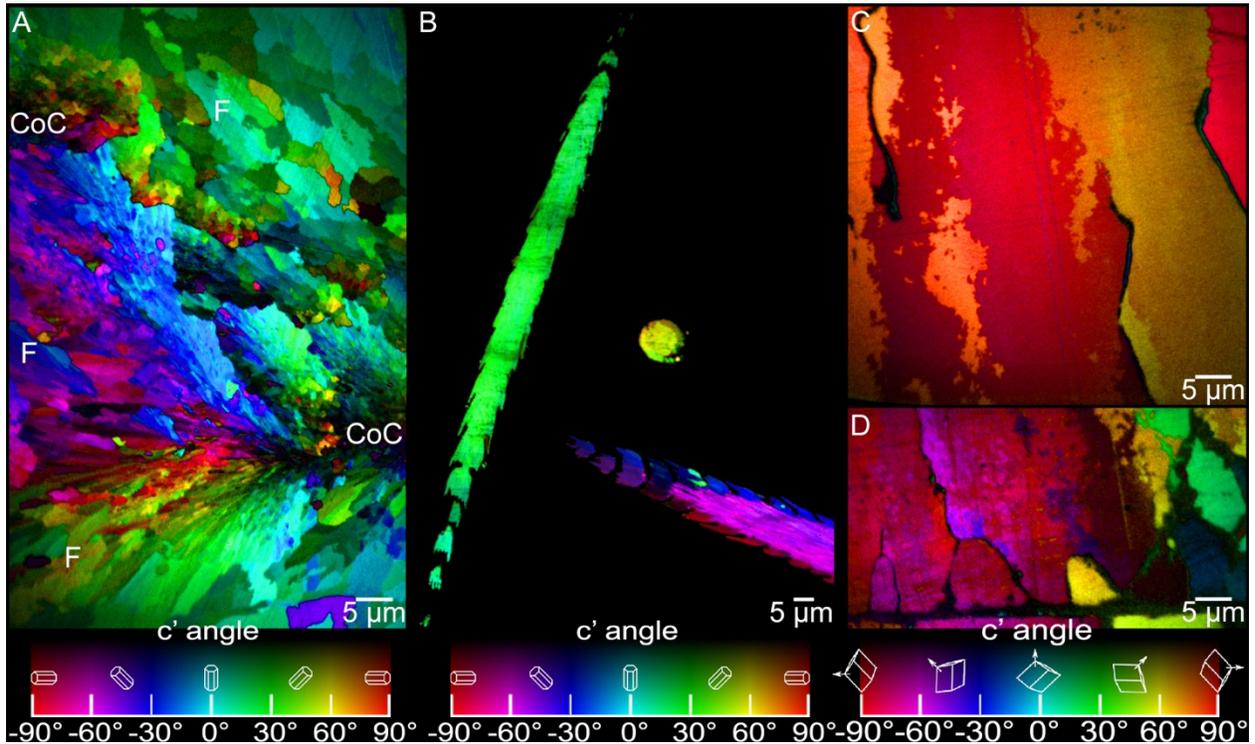

Figure 1. **Mesoscale crystal orientation in aragonite, vaterite, and calcite biominerals**.

A. Aragonite skeleton from *Stylophora pistillata* coral. Centers of calcification are between "CoC" labels, and acicular fibers "F" radiate from them.

B. Vaterite spicules from the tunicate ascidian *Hermania momus*.

C,D. Calcite prisms from the Tahitian pearl oyster *Pinctada margaritifera*. Black pixels are organic envelopes between calcite prisms.

All 3 biominerals show that adjacent crystallites have small (<30°) *c*-axis misorientations.

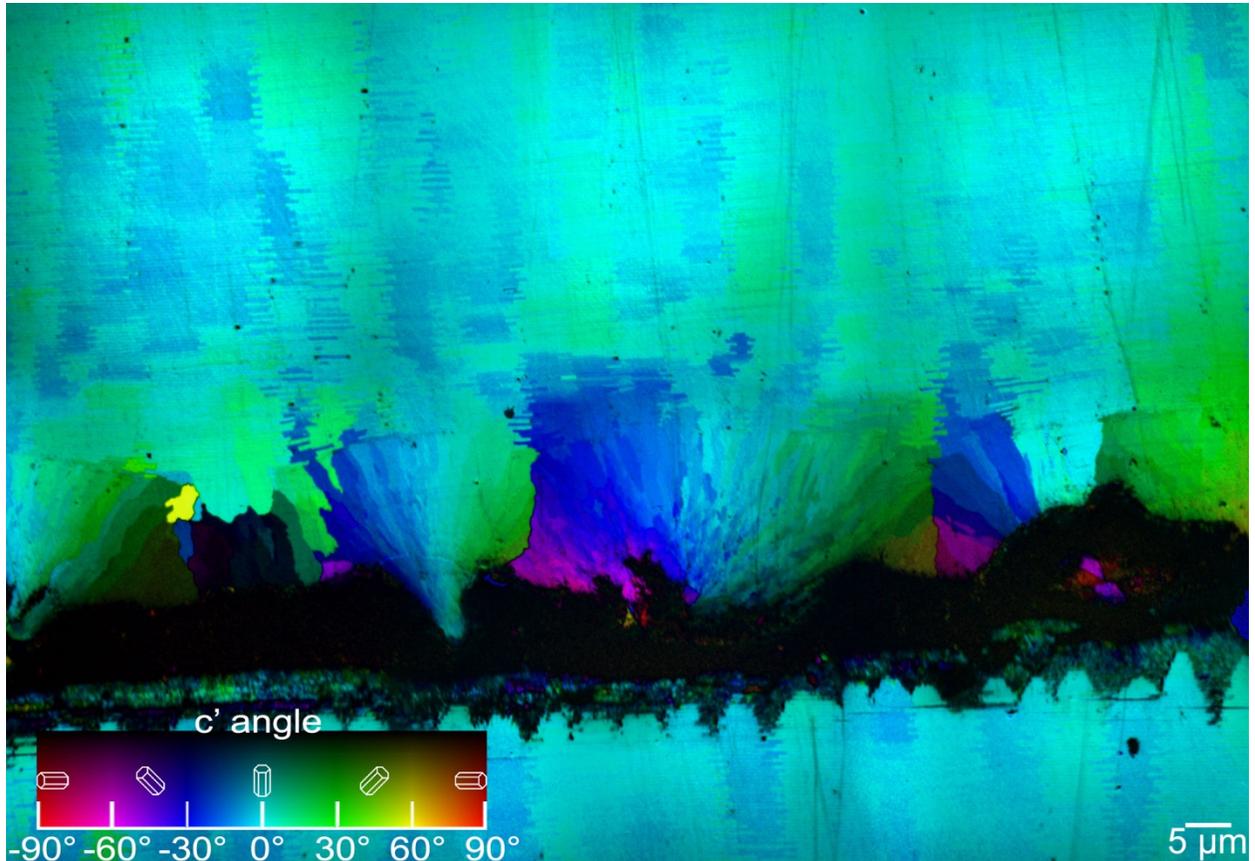

Figure 2. **Small misorientations in nacre start as spherulites**. Aragonite nacre from the California red abalone *Haliotis rufescens*. Nacre growth proceeds in the bottom-to-top direction in this cross-section. It is interrupted, an organic dark horizontal layer is formed, then spherulites nucleate, and nacre growth restarts. Notice that the *c*-axis orientations closer to the normal to nacre layers prevail (cyan, bluish- and greenish-cyan), and the others rapidly disappear, e.g. green, blue, magenta, and yellow.

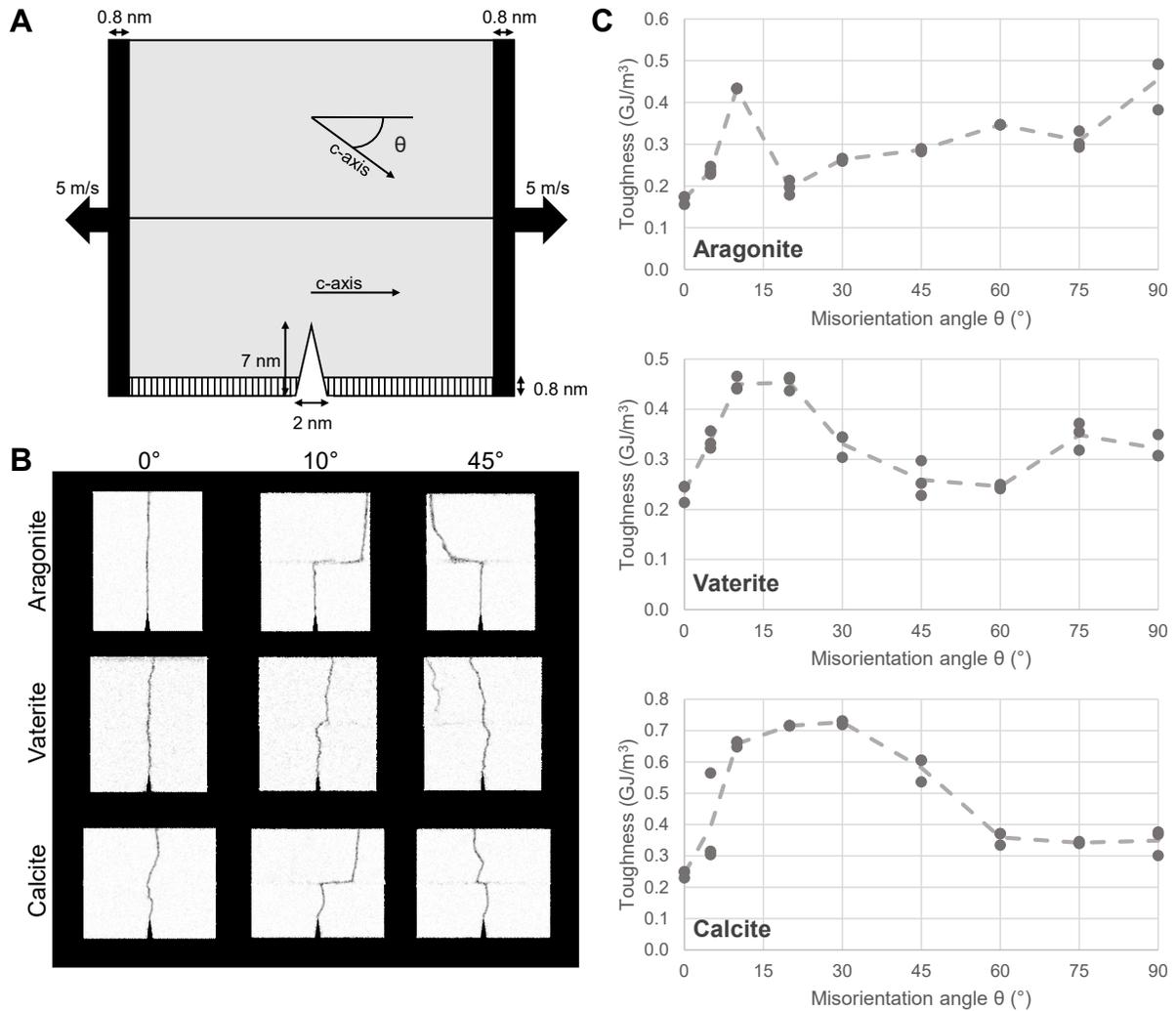

Figure 3: **Small-misorientation toughening.**
A. Aragonite, vaterite, and calcite were subject to mode I fracture. Bicrystal structures were created with the lower crystal *c*-axis parallel to the tensile direction and horizontal and the upper crystal *c*-axis rotated in the plane of the figure by a misorientation angle θ.
B. The introduction of a misoriented interface results in qualitative crack deflection, with a larger effect at 10° than 45°.
C. Fracture toughness calculated across a range of misorientations quantitatively shows that toughness peaks at small angles: approximately 10°, 20°, and 30° for aragonite, vaterite, and calcite, respectively.

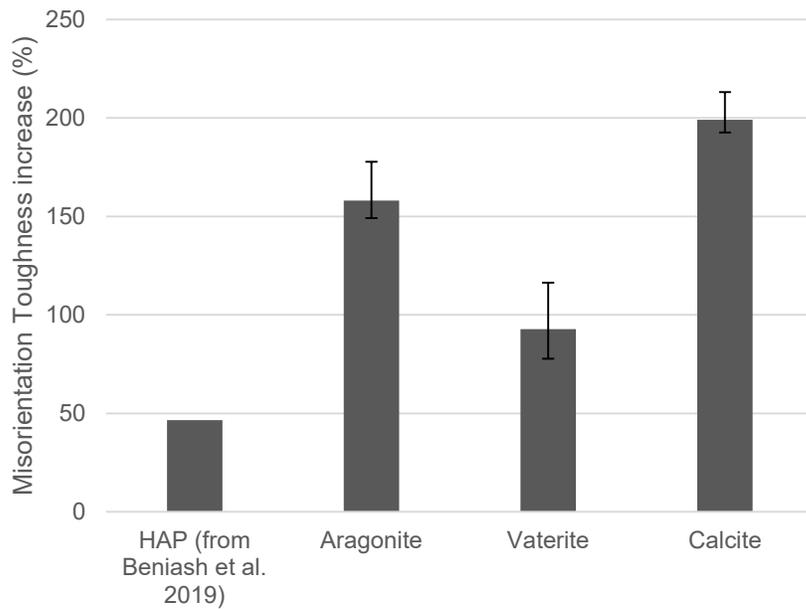

**Figure 4**: **Small misorientation makes bicrystals 2x-3x tougher.**
The percent increase in toughness from single crystal to small-angle misoriented bicrystals shows that peak fracture toughness values more than double for aragonite, vaterite, and calcite. Calcite experiences the greatest change in toughness, reaching values approximately triple the single crystal case. The small-angle misorientation effect previously identified in the literature for hydroxyapatite (HAP) is shown for comparison.

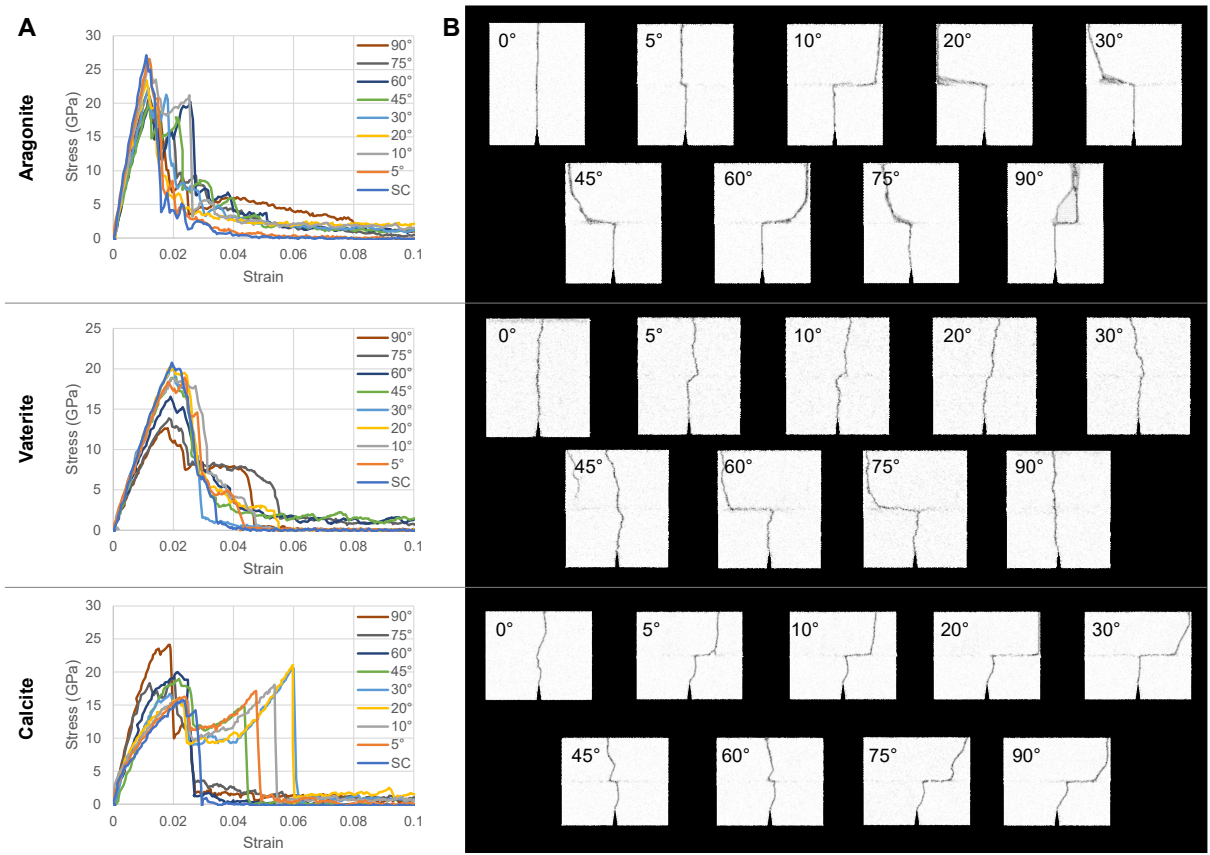

Figure S1: **Bicrystal fracture behavior.**
Fracture data for aragonite, vaterite, and calcite including:
A. Stress-strain curves.
B. Fracture paths obtained for single crystal and misorientation angles of 5°, 10°, 20°, 30°, 45°, 60°, 75°, and 90°.

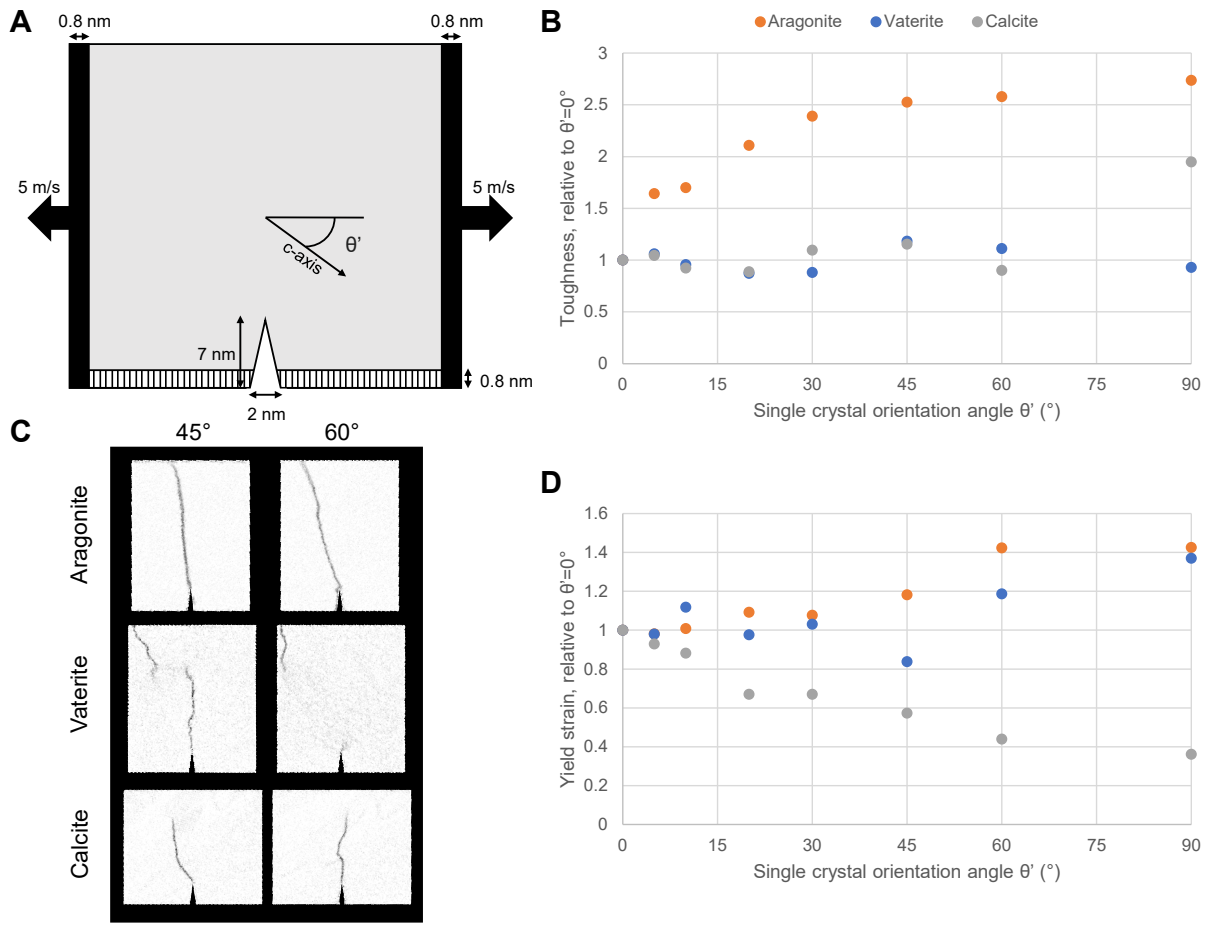

Figure S2: **Single crystal fracture behavior**.
At higher misorientation angles (θ>30°) bicrystal path tortuosity can increase due to mechanisms other than small-angle misorientation toughening. This behavior can be understood in terms of (A) single crystal fracture, with crystal orientations θ'. In the case of aragonite (B), this is because single crystal toughness increases as the single crystal is rotated to higher angles. For the case of vaterite (C), the single crystal exhibits secondary fracture from the upper plane when rotated to higher angles. And for the case of calcite (D), the yield strain decreases with rotation angle, favoring premature failure of the upper crystal as another mechanism of secondary fracture. These single crystal effects result in some path tortuosity in the bicrystal structures at higher misorientation angles, distinct from the small-angle misorientation effect.